# Nazarovets, S.A.

National University of Kyiv-Mohyla Academy Library, 2, Skovorody St., Kyiv, 04655, Ukraine, tel.: +380 44 425 6035, nazarovetssa@ukma.edu.ua

# WAR AND PEACE: THE PECULIARITIES OF UKRAINIAN-RUSSIAN SCIENTIFIC COOPERATION DYNAMICS AGAINST THE BACKGROUND OF RUSSIAN MILITARY AGGRESSION IN UKRAINE, IN 2014–2016

*The paper presents the results of bibliometric analysis of publications co-written by authors affiliated with Ukrainian and Russian institutions in 2007—2016, based on Scopus data. The survey results have shown that Ukrainian and Russian researchers continue carrying out joint research in major international projects. However, a decrease in the number of works published by Ukrainian and Russian research institutions in 2016 has testified to an adverse impact of Russia's military aggression on cooperation in the field of science. The findings are important for preparing the science development programs in Ukraine.*

*Keywords: scientific cooperation, publication activity, bibliometric analysis, Ukraine, and Russia*

Collaboration among researchers from different countries simplifies acquiring of new knowledge, expands opportunities for further use of research results, and facilitates the effective sharing of skills, competences, and resources. The Ukrainian science has directly suffered from the Crimea occupation by the armed forces of the Russian Federation in 2014 and the subsequent Russian armed aggression in the east of Ukraine. The employees of research institutions and higher educational establishments of Crimea and Donbass were forced to evacuate, while most equipment and materials for research remained in the occupied territories [1].

In addition, under conditions of undeclared war between Ukraine and Russia, the scientists of both countries, due to objective and subjective reasons, may not be interested in establishment, continuation, and development of further cooperation that would be mutually beneficial in peacetime. Therefore, the purpose of this paper is to study the impact of Russian military intervention in Ukraine on the joint scientific activities of Ukrainian and Russian researchers by means of bibliometric analysis of co-authored scholarly research publications issued by Ukrainian and Russian institutions in 2014—2016.

As a rule, the co-authored publications by researchers from different countries have more quotes than those issued by authors from one country [2]. Thus, the researchers are interested in publishing works in co-authorship with foreign partners, because it stimulates the growth of the readership and the world recognition of their contribution to the development of science. The results of previous studies indicate that Ukraine belongs to Russia's main scientific partners and, Russia is an important scientific partner of Ukraine. At the same time, for both countries this partnership is not the top-priority, as Ukrainian and Russian researchers work much more effectively with colleagues from other countries.

An analysis of the scientific publications of Russian researchers issued in international co-authorship in 1999—2008 has shown that Ukraine is the only out of all post-Soviet countries among the top twenty countries: it was ranked the 12th in 1999—2003 and 15th in 2004—2008; the share of joint publications of Ukrainian and Russian scholars in the total number of Russian publications co-authored with international colleagues made up 3.7%, in 1999—2003, and 3.8%, in 2004—2008 [5, 13—14]. The dynamics of main indicators of publication activity and the citation of Russian scientific publications in 1996—2010 have shown that the share of Russian publications co-authored with Ukrainian scientists accounted for 5.4%, in 2010 [6].

Also, upon the results of interviews with experts and science advisers to foreign embassies to Russia, as well as with representatives of leading Russian universities and research organizations, Ukraine has been listed among the most promising countries for the development of international R&D cooperation for the next 5—10 years [7].

So far, no general analysis of international cooperation with Ukrainian scientists involved has been made, and domestic bibliometric studies of academic collaborations have been limited to individual research areas. The scientometrics analysis of international cooperation of Ukraine in the socio-humanitarian field has shown that the most active Ukrainian researches collaborate with colleagues from the United States, Germany, and the United Kingdom; whereas, they have much fewer common publications with Russian socio-humanist researchers [3].

In the context of Ukrainian-Russian scientific cooperation it should be noted that many competitions for fundamental research projects and their further implementation have been conducted jointly by the State Fund for Fundamental Research of Ukraine (SFFR) and the Russian Fund for Fundamental Research (RFFR). The competitions were marked with a record-breaking number of applications, with 288 projects of Ukrainian and Russian researchers scoring a success [4].

The joint scientific publications of researchers affiliated in Ukraine and Russia were searched on April 13, 2017, in the Scopus abstract database that provides the most complete overview of the research results in various scientific disciplines. The two types of scientific papers, the articles and the conference proceedings, (ProceedingsPaper) for the period of 2007—2016 were considered. The search was initially conducted in the Affiliationcountry field that should contain the word "Ukraine" to indicate that the author was employed at a Ukrainian institution. Then, documents of the Ukrainian researchers made in cooperation with Russian colleagues (LIMIT-TO(AFFILCOUNTRY, "RussianFederation")) were selected. Also, an additional search was conducted for the selection of documents co-authored by researchers of Ukraine and Russia only.

For the period of 2007—2016, 7840 documents (6655 articles and 1185 proceeding papers) coauthored by the staff of Ukrainian and Russian institutions are found. The number of these joint works of Ukrainian and Russian scholars includes publications with other foreign colleagues en gaged, among which the largest share belongs to researchers from USA (1857 publications), France (1654 publications), Poland (1599 publications), Italy (1561 publications), and Great Britain (1539 publications). Only 3631 documents (3030 articles and 601 proceedings) are published by researchers from Ukrainian and Russian R&D institutions (Fig. 1).

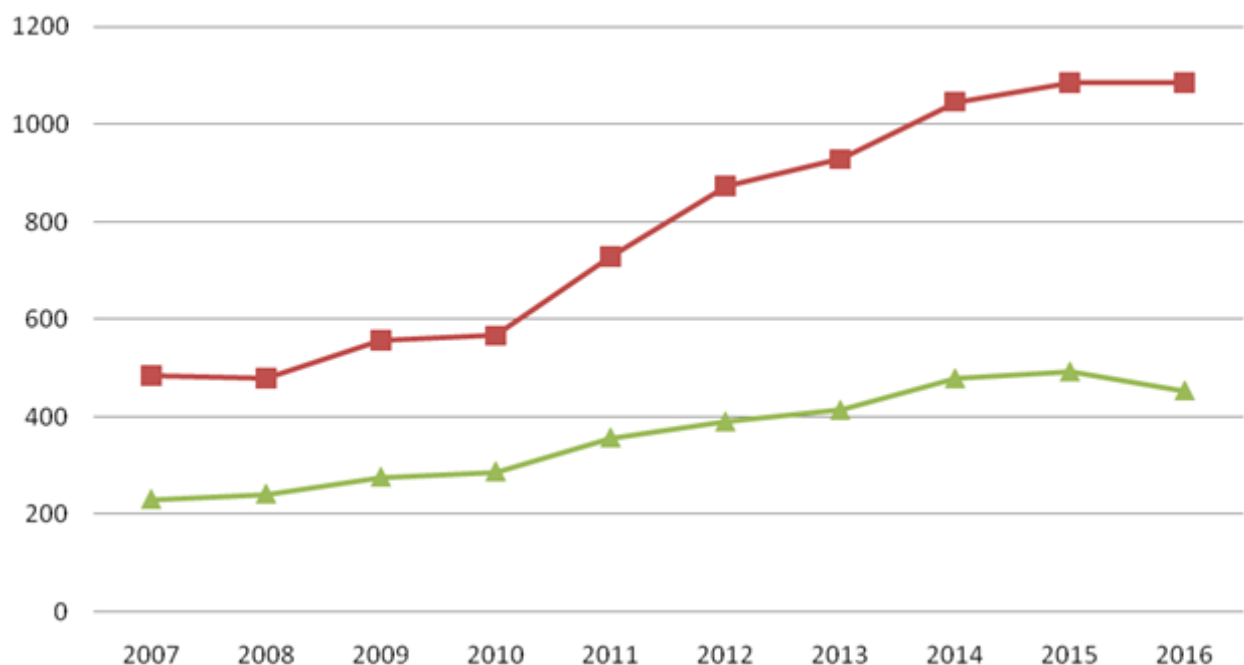

***Fig. 1.*** The number of joint documents co-authored by Ukrainian and Russian researchers with foreign colleagues engaged and the number of documents co-authored only by Ukrainian and Russian researchers, 2007—2016 based on Scopus data.

The most productive institutions in terms of the number of publications co-authored exclusively by Ukrainian and Russian researchers are as follows: the Russian Academy of Sciences (491 publications), the National Academy of Sciences of Ukraine (387 publications), the Lomonosov Moscow State University (331 publications), the Karazin Kharkiv National University (230 papers), and the Taras Shevchenko National University of Kyiv (218 publications). Thus, it can be stated that in 2007—2016 Ukrainian and Russian researchers cooperated both at the level of academic institutes and at the university level.

The analysis of documents co-authored by Ukrainian and Russian researchers has enabled to determine the most productive areas of Ukrainian-Russian cooperation, in accordance with the Scopus classification. Twelve subject areas have been selected, in which over 100 documents were published for the period of 2007—2016 (Fig. 2).

Ukrainian and Russian periodicals whose English-language versions belong to Springer Nature publisher prevail among the scientific journals that have published the largest number of documents based on the results of cooperation between Ukrainian and Russian researchers. It should be noted that during the last decade, the number of citations of these editions increased significantly, even if self-citations are not taken into account (Fig. 3). The largest number of documents was published in the following journals: Russian Journal of Inorganic Chemistry (88), Problems in Atomic Science and Technology (72), Bulletin of the Russian Academy of Sciences: Physics (60), Semiconductors (58), and Low Temperature Physics (57).

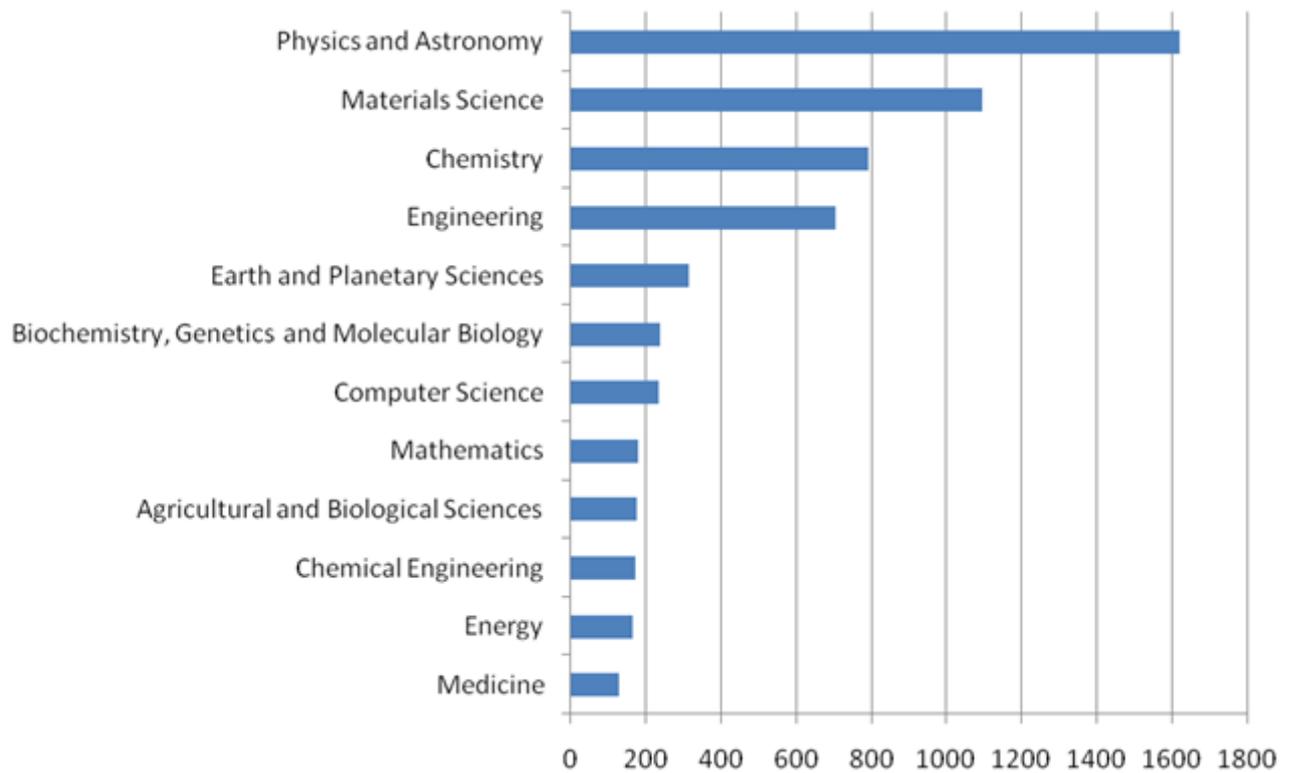

*Fig. 2.* The most productive fields of Ukrainian-Russian cooperation in 2007—2016, in accordance with Scopus classification

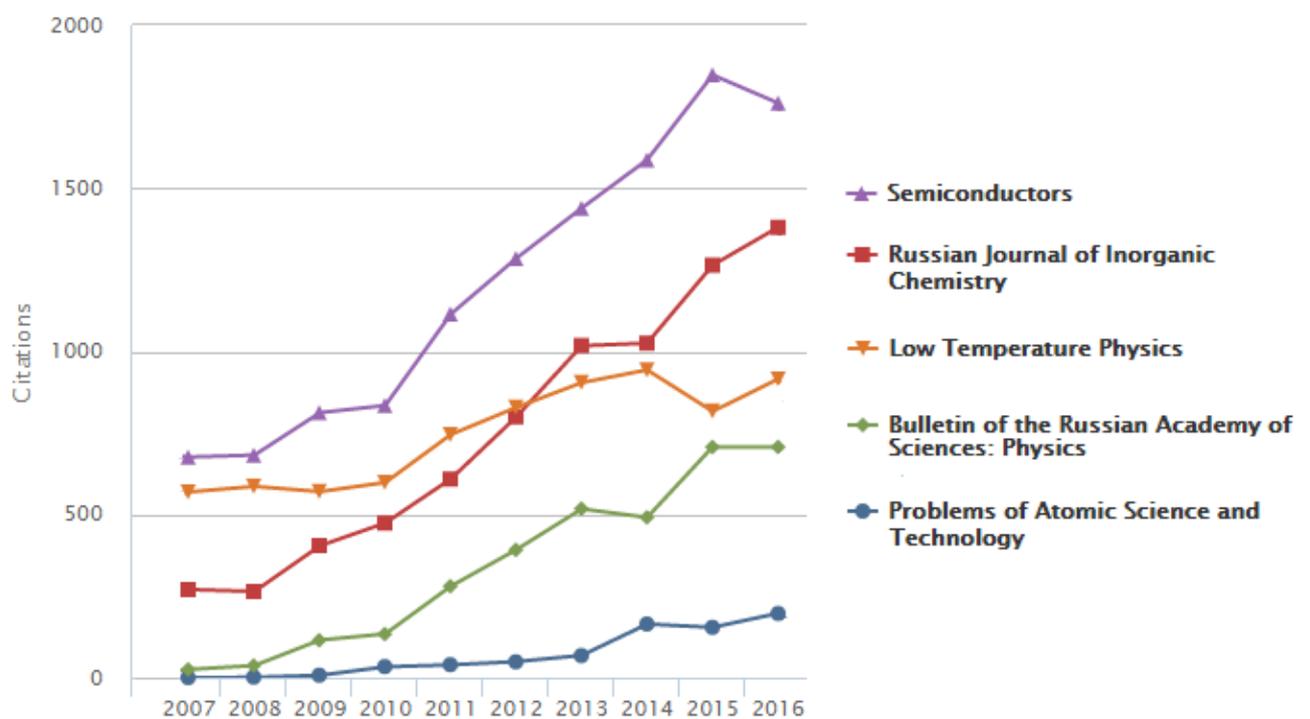

*Fig. 3.* Scientific journals that have published the largest number of documents upon the results of Ukrainian-Russian scientific cooperation in 2007—2016, based on Scopus data:

The obtained results show that the joint publishing activity of Ukrainian and Russian researchers with foreign co-authors engaged, has not significantly changed after the beginning of the hybrid war between Ukraine and Russia. Therefore, it can be concluded that Ukrainian and Russian scientists have not refused to participate in international scientific collaborations, even though they may engage researchers from the aggressor country.

It is clear that in 2015 and in 2014, the researchers published results of joint research completed before the beginning of armed conflict, therefore the data for these years may be unrepresentative. In this context, a slight decrease in the number of works co-authored by researchers from Ukrainian and Russian R&D institutions in 2016 could be interpreted as a negative impact of Russian armed aggression against Ukraine in 2014—2016 on the future prospects of Ukrainian-Russian scientific cooperation.

If such a trend really takes place in the Ukrainian-Russian scientific communication, then the downward dynamics in the number of works coauthored by Ukrainian and Russian researchers will continue in subsequent years, and, accordingly, the study of this problem will require further monitoring. However, this decrease in the number of joint publications in 2016 is not so meaningful as compared with previous years, and can be explained, in particular, by the loss of Ukraine's scientific infrastructure that remained in the occupied territory.

At the same time, the collected data on the joint publishing activity of Ukrainian and Russian scientists do not allow us to conclude that this cooperation was remarkably effective and successful. In recent years, Ukrainian and Russian researchers have been cooperating more closely with scientific partners from other countries. The previous statements of Ukrainian and Russian government officials on the crucial importance of Ukrainian-Russian scientific cooperation were declarative rather than factual. A significant part of the results of this cooperation has been published in the regional periodicals not included in the world abstract databases, and, accordingly, remains inaccessible to the world academic community. This fact has negative consequences for the development of science in Ukraine.

International cooperation of Ukrainian scientists requires a separate detailed study, including a thorough analysis of the factors that encourage Ukrainian researchers to work together with international partners, and an outlook of sources of funding for joint research in Ukraine's priority scientific disciplines. Similar results will be of great practical importance both for the scientists themselves and for the top officials of government institutions interested in developing international academic cooperation and in improving the quality of domestic research.